# CNN-based fully automatic mitral valve extraction using CT images and existence probability maps

Yukiteru Masuda*1, Ryo Ishikawa*1, Toru Tanaka*1, Gakuto Aoyama*2, Keitaro Kawashima*2, James V. Chapman*3, Masahiko Asami*4, Michael Huy Cuong Pham*5, Klaus Fuglsang Kofoed*5, Takuya Sakaguchi*2, Kiyohide Satoh*1

*1 Canon Inc., Tokyo, Japan
*2 Canon Medical Systems Corporation, Tochigi, Japan
*3 Canon Medical Informatics, Minnetonka, USA
*4 Division of Cardiology, Mitsui Memorial Hospital, Tokyo, Japan
*5 Department of Cardiology and Radiology, Copenhagen University Hospital - Rigshospitalet & Department of Clinical Medicine, Faculty of Health and Medical Sciences, University of Copenhagen, Copenhagen, Denmark

**Abstract**

Accurate extraction of mitral valve shape from clinical tomographic images acquired in patients has proven useful for planning surgical and interventional mitral valve treatments. However, manual extraction of the mitral valve shape is laborious, and the existing automatic extraction methods have not been sufficiently accurate. In this paper, we propose a fully automated method of extracting mitral valve shape from computed tomography (CT) images for the all phases of the cardiac cycle. This method extracts the mitral valve shape based on DenseNet using both the original CT image and the existence probability maps of the mitral valve area inferred by U-Net as input. A total of 1585 CT images from 204 patients with various cardiac diseases including mitral regurgitation (MR) were collected and manually annotated for mitral valve region. The proposed method was trained and evaluated by 10-fold cross validation using the collected data and was compared with the method without the existence probability maps.  The mean error of shape extraction error in the proposed method is 0.88 mm, which is an improvement of 0.32 mm compared with the method without the existence probability maps.

1. **Background**

   Severe secondary mitral regurgitation is associated with a mortality rate of 40% to 50% over a 3-year period and with fatal consequences for patients who do not receive appropriate treatment [1]. There are currently several types of interventional and surgical procedures available for the treatment of MR [2]. Accurate information on the morphology and dynamics of the mitral valve is critical in order to determine the treatment plan for patients with MR, and medical images such as transesophageal echocardiography (TEE) and contrast-enhanced computed tomography (CT) are often used to obtain morphologic and dynamic information pertaining to the mitral valve [3]. In order to obtain accurate morphologic and dynamic information on the mitral valve, it is necessary to extract the shape of the mitral valve from medical images, but manual extraction is very laborious and therefore impractical for routine clinical use. Many automatic mitral valve shape extraction methods have been proposed [4][5][7][8]. However, the mitral valve morphology is difficult to extract because of the complex saddle shape, the large individual difference in the shape and the thin and fluttering characteristics of the valve structure. These previously described methods are not accurate enough or require the user to enter a partial landmark input [4][5][7][8].

   Our contributions are as follows:
   - We present a novel fully automatic mitral valve extraction method from input to output for all phases of 4D CT images.
   - We suggest that the accuracy of mitral valve shape extraction is improved by using CT images and existence probability maps.

2. **Related work**

   Although many methods have been proposed for segmentation of the mitral valve region from CT and TEE imaging data, only a few methods automatically segment the mitral valve regions from images, and most of them require user operations to input landmarks or other information [4]. A fully automated mitral valve segmentation model using a physical mitral valve model applying patient-dependent parameters was proposed [5]. However, the performance of this method depends on the accuracy of extracting landmarks to estimate the deformation parameters of the physical mitral valve model, and the accuracy of the segmentation may be reduced if the landmarks are difficult to detect.

   In recent years, neural network technology such as U-Net has become widely used for medical image segmentation [6]. Neural network based-segmentation methods can segment the entire target region regardless of the accuracy of detection of the landmarks [7]. However, neural network based-segmentation methods are difficult to apply to organs with thin and fluttering structure such as mitral valve, because movement relative to volume is larger than in other organs and the individual difference in their movement and shape [8]. A method has been proposed to segment the mitral valve region from TEE images using an updated U-Net, but this method has a problem with the high number of false positives that resulted in the non-mitral valve region being incorporated into the mitral valve region segmentation, caused by a class imbalance in which the non-mitral valve region was overwhelmingly larger than the mitral valve region.

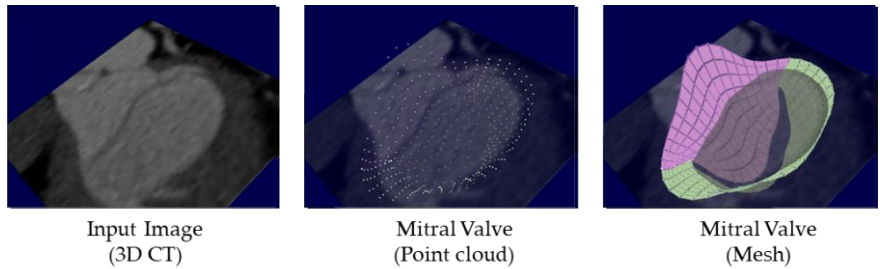

**Fig. 1** Input CT image and ground truth of Mitral Valve

3. **Materials and Methods**

3.1. Data Collection

We collected cardiac 3D- and 4D- CT imaging data from 204 patients at multiple medical facilities. These patients presented with various cardiac diseases including MR. The CT images were acquired with contrast using Aquilion ONE (Canon Medical Systems Corporation, Otawara, Tochigi, Japan) or SOMATOM Force (SIEMENS, Munich, Germany). The ECG gated cardiac CT sequences include 4–20 images per cardiac cycle, where each image contains 172–667 slices with 512 * 512 pixels. These data included a total of 1585 phase image data. The in-slice resolution is isotropic with 0.311–0.625 mm and a slice thickness of 0.5 to 0.75 mm.

The ground truth annotations were created by radiological technologists or general technologists under the guidance of expert physicians using in-house annotation tools. The ground truth annotation was defined as the leaflet position of the mitral valve, producing quadmesh coordinates of 19 x 9 for anterior leaflets and 25 x 9 for posterior leaflets. (Fig. 1).

3.2. Method overview

In this paper, we propose a method to perform a fully automatic extraction of the mitral valve shape with high accuracy. Figure 2 shows a flowchart for our method, which takes cardiac CT image including the mitral valve region as input, and outputs the mitral valve region represented by the quadmesh coordinates. Our method consists of three steps: First, perform a rough detection of the mitral valve region and crop a sub-region including mitral valve from the input CT image based on the detected region. Second, infer the existence probability maps of the mitral valve region for each voxel in the cropped image. Third, extract the mitral valve shape from cropped image with the probability maps.

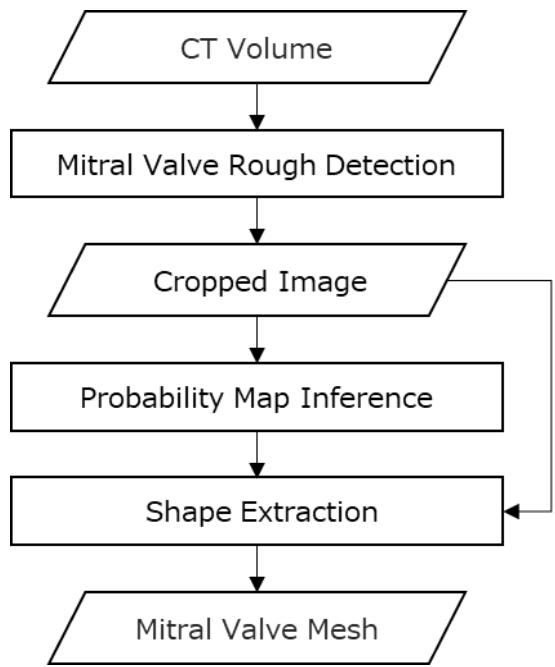

**Fig. 2** Flowchart of our method

3.3.1.  Mitral valve rough detection

The mitral valve region is roughly detected from the input CT image and a sub-region including the mitral valve is cropped from the input CT image based on the detected mitral valve region as a preprocessing for input to the neural network described later. To perform the rough detection of the mitral valve, the annular region of the mitral valve was extracted as a landmark. To detect the annular region of the mitral valve, we used the left ventricle and left atrium region masks, segmented using existing functionality provided by the Vitrea workstation (Canon Medical Systems Corporation, Japan). These left atrial and left ventricular region masks were dilated with 5 mm, and the region of the dilated masks boundary defined the annular region of mitral valve. The input CT image was cropped with a voxel size of 64 * 64 * 64 centered on the center of gravity of the extracted annulus region to include the entire extracted annulus region. The cropping position, orientation and scaling were corrected by the two points of the fibrous trigone and the position of the annular region of mitral valve opposed to it. These points are automatically detected by the VGG 16 deep learning model [9].

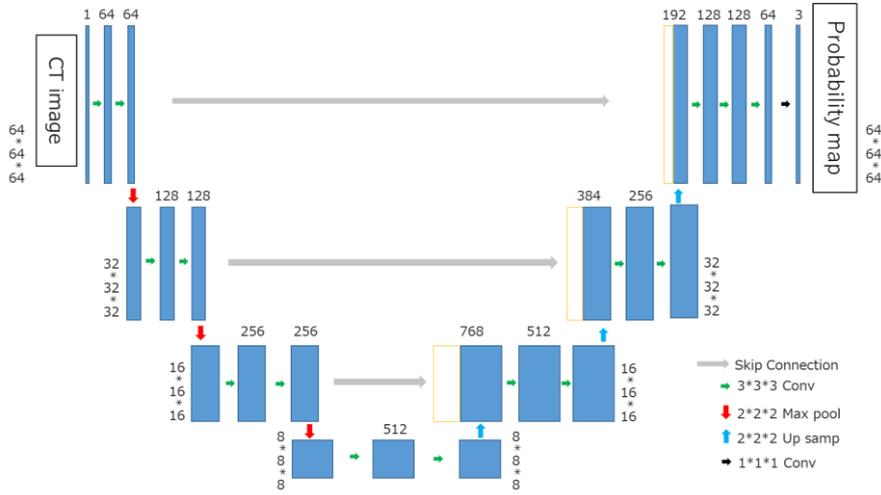

**Fig. 3** Probability map inference model architecture (U-Net)

In addition, each pixel value of cropped images was normalized in a range of 0 - 1.

3.3.2. Probability map inference

The existence probability map of the mitral valve region is inferred based on the 3D U-Net [10] with layer architecture shown in Figure 3. 3D U-Net consists of convolutional layer that performs convolutional processing to extract image features, pooling layer that reduces the image size so that more global image features can be acquired, and up sampling layer that re-expands the reduced image and returns it to the original image size. A skip connection structure is also provided for copying and adding a feature map of the same size in a down-sampling layer at the up-sampling layer. Each convolution layer applied 3 * 3 * 3 kernel and pooling layer apply 2 * 2 * 2 up sampling or down sampling. An activation function for Rectified Linear Unit (ReLU) was used. This 3D U-Net takes the cropped CT image as input and outputs a 3-channel probability maps of the following three classes 1. the anterior leaflet, 2. posterior leaflet, and 3. background (non-mitral valve region).

Multi class Tversky loss (Eq.1), which is updated Tversky [11], was used as the loss function. This loss function is useful for controlling over-extraction when the number of target voxels is much smaller than the number of non-target voxels. Multi class Tversky loss for each class *k* is defined as:

$$\text{Loss}_T(y_k, \hat{y}_k) = 1 - \frac{\sum_{i=0}^{N} y_{0ik}\hat{y}_{0ik}}{\sum_{i=0}^{N} y_{0ik}\hat{y}_{0ik} + \alpha_k \sum_{i=0}^{N} y_{1ik}\hat{y}_{0ik} + \beta_k \sum_{i=0}^{N} y_{0ik}\hat{y}_{1ik}}$$

(Eq.1)

$y_k$ means the ground truth of the class k and $\hat{y}_k$ means the inferred map of the class *k*. *i* corresponds to each pixel in the probability map. $y_{0ik}$ is a value where voxels corresponding to the class are 1 and voxels different from the class are 0, and vice versa for $y_{1ik}$. $\alpha_k$ and $\beta_k$ are coefficients of false positive and false negative, respectively, and are different values for each class. Referring to the parameter settings in original Tversky loss [11], $\alpha_k$ for $k_{ant}$ and $k_{pos}$ is 0.3, $\alpha_k$ for $k_{BG}$ is 0.7, $\beta_k$ for $k_{ant}$ and $k_{pos}$ is 0.7 and $\beta_k$ for $k_{BG}$ is 0.3. The $\text{Loss}_{\text{MCT}}(y,\hat{y})$, which is the sum of the $\text{Loss}_T(y_k, \hat{y}_k)$ of each class, is defined as:

$$\text{Loss}_{\text{MCT}}(y,\hat{y}) = a * \text{Loss}_T(y_{ant}, \hat{y}_{ant}) + b * \text{Loss}_T(y_{pos}, \hat{y}_{pos}) + c * \text{Loss}_T(y_{BG}, \hat{y}_{BG})$$

(Eq.2)

*a, b, c* are coefficients of respective $\text{Loss}_T$, and in our model *a* = 0.49, *b* = 0.49, and *c* = 0.02.

3.3.3. Shape extraction

The mitral valve shape is extracted from cropped image by DenseNet-121[12] updated to take 3 channels of 3D CT image and the existence probability maps as input. The detailed layer architecture of this model is shown in Table 1. This shape extraction model outputs quadmesh coordinates consisting of 19 * 9 points for the anterior leaflet and 25 * 9 points for the posterior leaflet. The difference in the number of points in each leaflet is due to the length of each leaflet's annulus. The mean squared error at each point was used as the loss function. To prevent overlearning due to regularization, we use Mish[13] as activation function.

4. **Experiments**

4.1. Training setup

All training was performed using Tensorflow 2.4 as the backend on a workstation with NVIDIA GeForce RTX 3090 24 GB GPU. The initial learning rate of the model started from 0.00002 for the probability map inference and 0.002 for the shape extraction, and was divided by 10 when loss did not improve for 5 epochs. ADAM[14] optimizer was used for both models.

4.2. Learning data preprocessing

| Layers | Output Size | Kernel , stride |
|---|---|---|
| Convolution | 32 * 32 * 32 | 7 * 7 * 7 conv, stride 2 |
| Pooling | 16 * 16 * 16 | 3 * 3 * 3 Maxpooling, stride 2 |
| Dense Block (1) | 16 * 16 * 16 | $\begin{pmatrix} 1*1\ \text{conv} \\ 3*3\ \text{conv} \end{pmatrix} * 6$ |
| Transition Layer (1) | 16 * 16 *16 | 1 * 1 conv |
|  | 8 * 8 * 8 | 2 * 2 average pool, stride 2 |
| Dense Block (2) | 8 * 8 * 8 | $\begin{pmatrix} 1*1\ \text{conv} \\ 3*3\ \text{conv} \end{pmatrix} * 12$ |
| Transition Layer (2) | 8 * 8 * 8 | 1 * 1 conv |
|  | 4 * 4 * 4 | 2 * 2 average pool, stride 2 |
| Dense Block (3) | 4 * 4 * 4 | $\begin{pmatrix} 1*1\ \text{conv} \\ 3*3\ \text{conv} \end{pmatrix} * 24$ |
| Transition Layer (3) | 4 * 4 * 4 | 1 * 1 conv |
|  | 2 * 2 * 2 | 2 * 2 average pool, stride 2 |
| Dense Block (4) | 2 * 2 * 2 | $\begin{pmatrix} 1*1\ \text{conv} \\ 3*3\ \text{conv} \end{pmatrix} * 16$ |
| Output Layer | 1024 | 2 * 2 * 2 Global Average Pool |
|  | 1188 | 1188D fully-connected |

**Table. 1** Shape extraction model architecture (DenseNet)

The 1585 cardiac CT image data obtained from 204 patients were cropped and normalized to 3D images with 64 * 64 * 64 voxels by the method described in 3.3.1. Data augmentations including pixel value translation, rotation, scaling, and smoothing were performed on these data, and 22 augmented images from a single image were produced. All CT images were randomly split 8:1:1 for training, validation, and test steps, respectively, for 10-fold cross validation.

4.3. Evaluation

4.3.1 Evaluation Measures

In this experiment, the chamfer distance (CD) was used as a metric for the extraction accuracy of the quadmesh coordinates extracted as the mitral valve shape. Define the CD as follows:

$$CD_{(S(S), S(G), sS, sG)} = \frac{1}{q}\sum_{n=1}^{q}\left\{\frac{1}{|S(S_n)| + |S(G_n)|}\left(\sum_{sS_n \in S(S_n)} dm(sS_n, S(G_n)) + \sum_{sG_n \in S(G_n)} dm(sG_n, S(S_n))\right)\right\}$$

(Eq.3)

$q$ means the number of phases in each patient $S(S_n)$ and $S(G_n)$ means surface of the inferred mesh, and surface of the ground truth mesh, respectively. $sS_n$ and $sG_n$ mean the points that make up the mesh and the ground truth, respectively. $dm$ mean the minimum Euclidean distance between a point and a surface.

We also calculated the Hausdorff distance (HD) to measure the point of greatest error in each mesh. Define the HD as follows:

$$HD_{(S(S), S(G), sS, sG)} = \frac{1}{2}\left\{\max_{sS \in S(S)}\{d_m(sS, S(G))\} + \max_{sG \in S(G)}\{d_m(sG, S(S))\}\right\}$$

(Eq.4)

To prevent confusion between the anterior and posterior leaflets, CD and HD are calculated separately for the anterior and posterior leaflets, and the average value is used as the overall value. The time required for shape extraction was also measured. And extracted mitral valve shapes were visually examined for the collapse of the shape at all phases.

4.3.2 Evaluation conditions

To measure the effect of adopting the probability maps to the shape extraction model, the CD and HD was compared with and without the probability maps respectively. In the shape extraction model without a probability maps, the input was reduced from 3 channels to 1 channel, and learning and inference were performed using only 3D CT images as input.

In order to examine whether the extraction of the mitral valve was possible even in MR cases in which the movement of the mitral valve deviated from normal, CD and HD were calculated and compared by dividing into normal cases and MR cases. In addition, the mitral valve was evaluated separately for the anterior and posterior leaflets to examine the mitral valve extraction performance of the model in more detail.

4.3.3   Statistical Analysis

For each evaluation measurement value, the mean and standard error (SE) were calculated for each patient. In addition, in order to examine whether the accuracy was significantly improved by adopting the probability maps, the results with and without the probability maps were analyzed by paired t-test. In this analysis, Bonferroni correction was used to adjust for multiple validation endpoint in the measurement items, and the significance level were considered if $p < 0.0083$. SPSS (IBM® SPSS® Statistics, version 26.0, IBM, NY) was used for all statistical analysis.

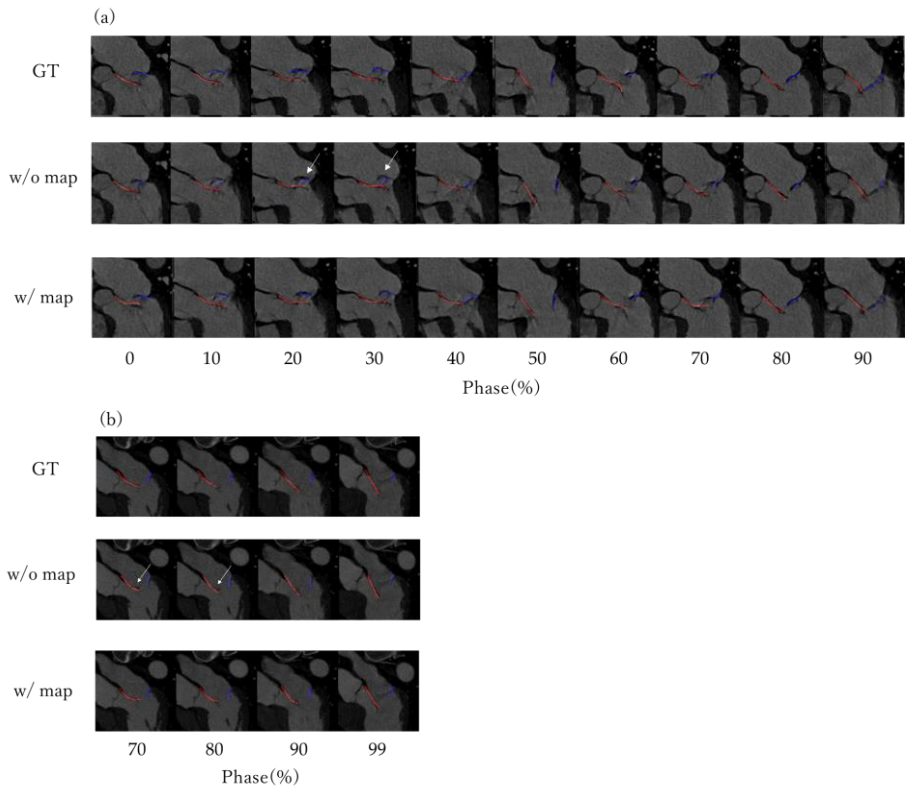

**Fig. 4** Comparison of inferred meshes cross-section with and without probability maps. The mesh consists of grid points connected by lines, with the anterior leaflet in red and the posterior leaflet in blue. Arrows in the figure indicate areas where errors in the mesh shape are large. (a) MR case. (b) Normal case.

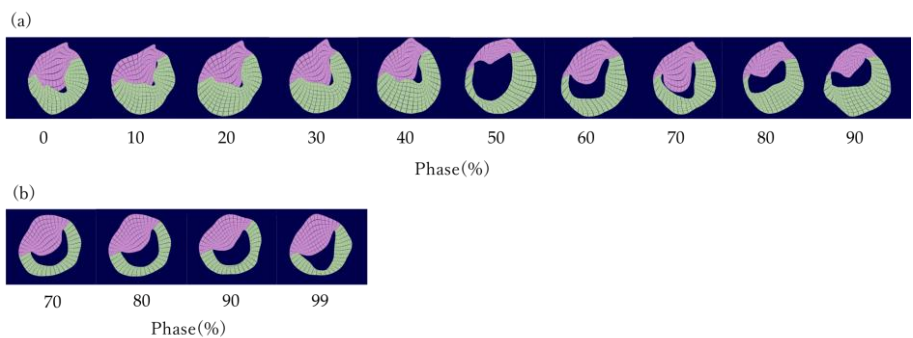

**Fig. 5** Mitral valve mesh extracted by shape extraction model with probability maps. The anterior leaflet of the mitral valve is shown in pink and the posterior leaflet in green. (a) MR case. (b) Normal case.

5. **Results**

In the proposed method, extraction of the mitral valve shape was successful in all cases and the time required for shape extraction about 7 sec per phase.

Figure 4 shows an image in which the estimated mesh in all phases of the shape extraction with probability maps and without probability maps is overlaid on the original CT image. In some cases, the shape extraction without probability maps cannot recognize the edge of the valve, but in the with probability maps model, the recognition accuracy of the edge of the valve is improved. Figure 5 shows the 3D mesh shape at all phases of the mitral valve extracted by a shape extraction model with a probability maps. Table 2 shows the evaluation results of the shape extraction model with probability maps and the model without probability maps. The mean CD of the shape extraction model with probability maps is 0.88 mm, and the CD of the model without probability maps model is 1.20 mm. This result shows that the accuracy of the entire final mesh is improved by adding probability maps to the shape extraction model input. In addition, the Hausdorff distance (HD) is 4.19 mm for the shape extraction model with probability maps and 4.47 mm for the model without probability maps, indicating that the maximum distance error can also be reduced. Comparing the normal case and the MR case, the average improvement was 0.33 mm in the normal case and 0.31 mm in the MR case by adopting the shape extraction model with probability maps, indicating that the improvement effect was obtained in both cases. Table 3 shows the shape extraction errors of the anterior and posterior cusps, and it can be seen that the detection accuracy of the posterior cusps was particularly improved by adopting the probability maps.

6. **Discussion**

We propose a fully automatic mitral valve shape extraction method, which uses mitral valve probability maps as an input to the shape extraction model. Our proposed method showed that the average CD in all cases was 0.88 mm. Although a direct comparison is not possible due to the different data sets used for learning and evaluation, the accuracy of our method was superior compared with previous method in CD [5].

Mitral valve shape extraction was highly accurate in most patients, but extraction accuracy was significantly poor in 2 of 204 patients with or without probability maps. The reason for the poor accuracy of these images is thought to be that these images failed during the rough detection step of preprocessing the mitral valve, and the cropping of

|  | All | | Normal | | MR | |
|---|---|---|---|---|---|---|
|  | CD(mm) | HD(mm) | CD(mm) | HD(mm) | CD(mm) | HD(mm) |
| DenseNet | 1.20 ±0.033 | 4.47 ±0.081 | 1.10 ±0.042 | 4.12 ±0.090 | 1.44 ±0.037 | 5.30 ±0.113 |
| DenseNet (with map) | 0.88 ±0.031 | 4.19 ±0.084 | 0.79 ±0.040 | 3.85 ±0.094 | 1.11 ±0.035 | 5.00 ±0.124 |
| p-value | < 0.001* | < 0.001* | < 0.001* | < 0.001* | < 0.001* | < 0.001* |

**Table. 2** Comparison of the two models in the normal and MR cases.

Data are mean ± standard error for cases.

* showed *p* < 0.0083.

|  | All | | Anterior | | Posterior | |
|---|---|---|---|---|---|---|
|  | CD(mm) | HD(mm) | CD(mm) | HD(mm) | CD(mm) | HD(mm) |
| DenseNet | 1.20 ±0.033 | 4.47 ±0.081 | 1.08 ±0.023 | 4.50 ±0.078 | 1.29 ±0.046 | 4.45 ±0.099 |
| DenseNet (with map) | 0.88 ±0.031 | 4.19 ±0.084 | 0.81 ±0.020 | 4.32 ±0.083 | 0.94 ±0.046 | 4.09 ±0.104 |
| p-value | < 0.001* | < 0.001* | < 0.001* | < 0.001* | < 0.001* | < 0.001* |

**Table. 3** Comparison of the two models in the mitral valves by site.

Data are mean ± standard error for cases.

* showed *p* < 0.0083.

the images used as input for shape extraction was greatly shifted. For these patients, when inference was made by manually cropping the image, the CD was improved as for other patients. Mitral valve shape extraction with probability maps improves extraction accuracy in both normal patients and MR patients compared with extraction without probability maps. It is considered that our method was able to extract the features of the mitral valve shape more accurately by catching each of the local features by DenseNet and the global features by U-Net, and compensating for the missing parts in each model by stacking them.

Table 3 shows that the accuracy of extraction of posterior leaflet was lower than that of the anterior leaflet. This may be because the posterior leaflet is often located near the heart

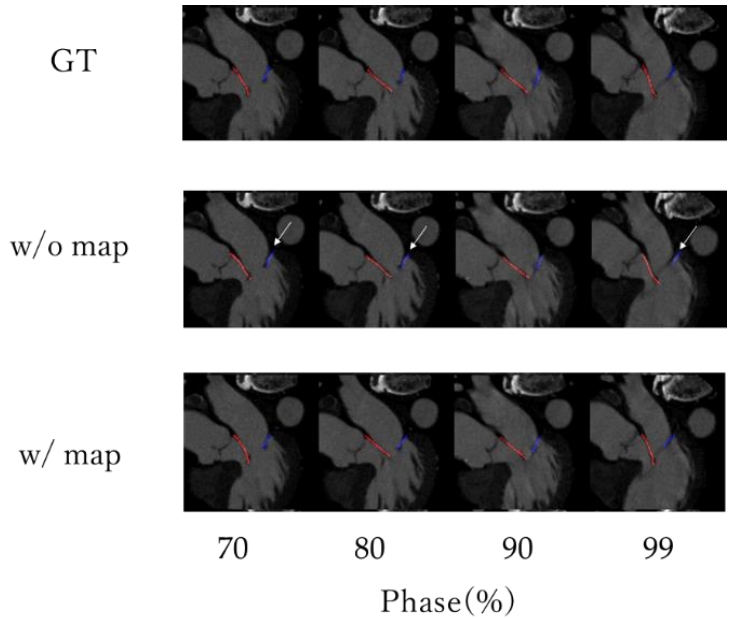

**Fig. 6** Comparison of inferred meshes cross-section with and without probability maps with greater improvement at the posterior leaflet.

The mesh consists of grid points connected by lines, with the anterior leaflet in red and the posterior leaflet in blue. Arrows in the figure indicate areas where errors in the mesh shape are large.

wall during the ED phase, making it difficult to separate the heart wall from the posterior leaflet. However, as shown in Figure 6 the accuracy of the extraction of the posterior leaflet can be improved by introducing the probability maps even in an image in which the extraction of the posterior leaflet fails without the probability maps. By introducing probability maps, not only local features but also global features can be considered, so the posterior leaflet, which is difficult to separate from the heart wall in the conventional model, can be extracted accurately.

The development of an automatic mitral valve measurement application based on the proposed method is expected to enable physicians to obtain mitral valve morphology information more quickly without inter-user variability for treatment planning. In addition, the quadmesh extraction of the mitral valve can also be used as input data for other techniques to support treatment planning, such as simulating valve structure and fluid flow.

As a limitation, the correlation between the image quality of the input image and accuracy has not been investigated. In addition, when considering clinical use, there is a possibility that it may be affected by the constraints of imaging conditions. The performance of our method has not been directly compared to the performance reported in previous studies using the same CT datasets. It has not been verified that the mitral valve shape extracted by this method is of acceptable accuracy for clinical use.

In future work, we will also examine whether the anatomy calculated from the mitral valve shape extracted by this method can contribute to clinical diagnosis and intervention decisions.

**7. Conclusion**

We developed a fully automatic extraction method of the mitral valve shape from CT images. We show that the extraction accuracy of the mitral valve shape was improved by using CT images and existence probability maps.

**References**

1. Lavall D, Hagendorff A, Schirmer SH, Böhm M, Borger MA, Laufs U. "Mitral valve interventions in heart failure." (2018) ESC Heart Fail. Aug;5(4):552-561. https://doi.org/10.1002/ehf2.12287
2. Catherine M Otto, Rick A Nishimura, Robert O Bonow, Blase A Carabello, John P Erwin 3rd, Federico Gentile, Hani Jneid, Eric V Krieger, Michael Mack, Christopher McLeod, Patrick T O'Gara, Vera H Rigolin, Thoralf M Sundt 3rd, Annemarie Thompson, Christopher Toly. "2020 ACC/AHA Guideline for the Management of Patients With Valvular Heart Disease: A Report of the American College of Cardiology/American Heart Association Joint Committee on Clinical Practice Guidelines" (2021) J Am Coll Cardiol. Feb, 77 (4) e25–e197. https://doi.org/10.1016/j.jacc.2020.11.035
3. Garcia-Sayan, Enrique, Tiffany Chen, and Omar Kamaal Khalique. "Multimodality Cardiac Imaging for Procedural Planning and Guidance of Transcatheter Mitral Valve Replacement and Mitral Paravalvular Leak Closure." (2021) Frontiers in Cardiovascular Medicine 8: 44. https://doi.org/10.3389/fcvm.2021.582925
4. Tiwari, Abhishek, and Kedar A. Patwardhan. "Mitral valve annulus localization in 3D echocardiography." (2016) Annu Int Conf IEEE Eng Med Biol Soc. Aug;2016:1087-1090. https://doi.org/10.1109/embc.2016.7590892
5. Razvan Ioan Ionasec, Ingmar Voigt, Bogdan Georgescu, Yang Wang, Helene Houle, Fernando Vega-Higuera, Nassir Navab, Dorin Comaniciu. "Patient-Specific Modeling and Quantification of the Aortic and Mitral Valves from 4D Cardiac CT and TEE" (2010) *IEEE Trans Med Imaging*. Sep, 29(9):1636-51. https://doi.org/10.1109/tmi.2010.2048756
6. Olaf Ronneberger, Philipp Fischer, Thomas Brox. "U-Net: Convolutional Networks for

Biomedical Image Segmentation" (2015) arXiv:1505.04597. https://doi.org/10.48550/arXiv.1505.04597

7. Borge Solli Andreassen, Federico Veronesi, Olivier Gerard, Anne H Schistad Solberg, Eigil Samset. "Mitral annulus segmentation using deep learning in 3-D transesophageal echocardiography." (2019) IEEE J Biomed Health Inform. Apr;24(4):994-1003. https://doi.org/10.1109/jbhi.2019.2959430

8. Eva Costa, Nelson Martins, Malik Saad Sultan, Diana Veiga, Manuel Ferreira, Sandra Mattos, Miguel Coimbra. "Mitral Valve Leaflets Segmentation in Echocardiography using Convolutional Neural Networks" (2019) *IEEE 6th Portuguese Meeting on Bioengineering (ENBENG)*, pp. 1-4. https://doi.org/10.1109/ENBENG.2019.8692573

9. Karen Simonyan, Andrew Zisserman "Very Deep Convolutional Networks for Large-Scale Image Recognition" (2015) arXiv:1409.1556 https://doi.org/10.48550/arXiv.1409.1556

10. Özgün Çiçek, Ahmed Abdulkadir, Soeren S. Lienkamp, Thomas Brox, Olaf Ronneberger. "3D U-Net: Learning Dense Volumetric Segmentation from Sparse Annotation" (2016) arXiv:1606.06650 https://doi.org/10.48550/arXiv.1606.06650

11. Seyed Sadegh Mohseni Salehi, Deniz Erdogmus, Ali Gholipour. "Tversky loss function for image segmentation using 3D fully convolutional deep networks" (2017) arXiv:1706.05721 https://doi.org/10.48550/arXiv.1706.05721

12. Gao Huang, Zhuang Liu, Laurens van der Maaten, Kilian Q. Weinberger "Densely Connected Convolutional Networks" (2016) arXiv:1608.06993 https://doi.org/10.48550/arXiv.1608.06993

13. Diganta Misra. "Mish: A Self Regularized Non-Monotonic Activation Function" (2019) arXiv:1908.08681 https://doi.org/10.48550/arXiv.1908.08681

14. Diederik P. Kingma, Jimmy Ba. "Adam: A Method for Stochastic Optimization" (2014) arXiv:1412.6980 https://doi.org/10.48550/arXiv.1412.6980

**Statements and Declarations**

**Funding**

This study was funded by Canon Medical Systems Corporation and Canon Inc.

**Ethical approval**

All procedures performed in human studies were in accordance with the ethical standards of the Institutional Research Committee and the 1964 Declaration of Helsinki, as amended or equivalent. Formal consent is not required for this type of study.

**Informed consent**

Obtaining informed consent was waived by the institutional review board because patient data were collected retrospectively in this study.

**Conflict of interest**

Y. Masuda, R. Ishikawa, T. Tanaka and K. Satoh are Canon Inc employees.

G. Aoyama and K. Kawashima are Canon Medical Systems Corporation employees.

J. Chapman is a Canon Medical Informatics employee.

K. F. Kofoed has received research grants from Canon Medical Systems Corporation and AP Møller og hustru Chastine McKinney Møllers Fond.

M. Asami has received remuneration for advice from Canon Medical Systems Corporation.

The other authors declare no conflict of interest.

**Author contributions**

study conception and design, Y.M., R.I., T.T., and G.A. ; software, Y.M., R.I., and T.T.; validation, Y.M. and R.I.; investigation, Y.M., R.I., T.T., G.A., K.K., and J.C.; data collection, G.A., K.K., M.H.C.P., and K.F.K. writing—The first draft of the manuscript, Y.M.; writing— review and editing, R.I., G.A., T.T., K.S., and T.S.,; supervision : M.A., J.C., M.H.C.P., K.F.K., K.S., T.S.; project administration, R.I., G.A., T.S., and K.S.,; All authors read and approved the final manuscript.